\documentclass{kapproc}
\usepackage{procps} 
\usepackage[dvips]{graphicx}
\upperandlowercase
{\normallatexbib

\begin{document}
\newcommand{\lsim}{\stackrel{\scriptstyle <}{\phantom{}_{\sim}}}
\newcommand{\gsim}{\stackrel{\scriptstyle >}{\phantom{}_{\sim}}}

\articletitle{Thermal Color-superconducting\\
Fluctuations
in Dense
Quark Matter}


\author{D.N. Voskresensky}
\affil{Gesellschaft f\"ur Schwerionenforschung mbH, Planckstr. 1,
64291 Darmstadt, Germany;\\
 Moscow Institute for Physics and Engineering, 
Kashirskoe sh. 31, Moscow 115409, Russia}
\email{D.Voskresensky@gsi.de}


\begin{abstract}
Thermal fluctuations of the color superconducting order parameter in dense
quark matter are investigated in terms
of the phenomenological Ginzburg - Landau approach. Our estimates
show that fluctuations of the di-quark gap may strongly affect
some of thermodynamic quantities even far below and above the
critical temperature.  If the critical temperature $T_c$ of the di-quark
phase transition were rather high ($\gsim (50\div 70)$~MeV)
one could expect a manifestation
of fluctuations of the di-quark gap 
in the course of heavy ion
collisions (above $T_c$). For $T_c \sim 50~$MeV color superconducting
fluctuations may also affect an initial stage of the hybrid star evolution.
\end{abstract}


\section*{Introduction}

The quark-quark interaction in the color antitriplet channel is
attractive driving the pairing, cf. \cite{bl}. The problem has
been re-investigated in a series of papers following Refs. \cite
{arw98,r+98}, see review \cite{RW} and Refs therein. The
attraction comes from the one-gluon exchange, or from a
nonperturbative 4-point interaction motivated by instantons
\cite{dfl}, or nonperturbative gluon propagators \cite{br}. The
zero-temperature pairing gap $\Delta$ was predicted to be $\sim
(20\div 200)$~MeV for the quark
chemical potentials $\mu_q \sim (300\div 500)$~MeV. 
In the
standard BCS theory, cf. \cite{RW}, the critical 
temperature is estimated as $T_c \simeq 0.57 \Delta$.

One expects the di-quark condensate to dominate the physics at
densities beyond the deconfinement/chiral restoration transition
and below the critical temperature. Various phases are possible.
E.g., the so called 2-color superconductivity (2SC) phase allows
for unpaired quarks of one color. There may also exist a
color-flavor locked (CFL) phase \cite{arw99} for not too large value of the
strange quark mass $m_s$, for $2\Delta >m_s^2 /\mu_q$,
cf.  \cite{abr99}, where the color
superconductivity (CSC) is complete in the sense that the di-quark
condensation produces a gap for quarks of all three colors and
flavors. The values of the gap 
are of the same order of magnitude for 2SC and CFL phases,
whereas 
relations between critical temperature and the gap
might be different, $T_c \simeq 0.57 \Delta$ for 2SC and 
$T_c \simeq 0.7 \Delta$ for CFL phase \cite{Sh00}.
There are also another possibilities, e.g.,
of the pairing in spin-one channel, for which the pairing gap proves to be 
small $\Delta \lsim 1$~MeV, see \cite{Sh00}.

The high-density phases of QCD at low temperatures may exist in
interiors of most massive neutron stars (so called hybrid stars
containing the quark core and the hadron shell) affecting the cooling,
rotation and magnetic field, cf. \cite{bgv,bss,IB}. It is also
possible to ask whether CSC is relevant for the terrestrial experiments?
To produce a color superconducting matter  in the laboratory one
would need to cook a dense and not too hot baryon enriched matter.
The nuclear matter prepared in heavy ion collisions at SIS,
AGS, SPS or RHIC in all the cases has presumably not sufficiently high
baryon density and, on the other hand, the matter prepared at SPS
or RHIC  is feasibly  too hot in order to expect a manifestation
of the CSC.
The most relevant is probably the GSI ``Compressed baryon matter'' 
future heavy ion collision
facility which may cook enough dense and not too hot 
state. Baryon densities up to ten normal nuclear matter
density ($\lsim 10\rho_0$) at temperatures $T\lsim 170$~MeV are
expected to be reached at an initial collision stage. It is
supposed that the system is in the quark-gluon plasma state at
such conditions. Then the system expands and cools down. In this
process the temperature decreases  up to $T\sim 140$~MeV at still
rather high baryon density at an intermediate collision stage.
Although the temperatures at relevant densities are most likely
larger than the critical temperature $T_c$ of CSC, one may rise
the question on a possible manifestation of the precursor
phenomena of the CSC phase transition, if $T_c$ is rather high
($T_c \gsim (50\div 70)$~MeV). Recently, ref. \cite{KKKN02}
considered such a possibility within the Nambu-Jona-Lasinio  model
and demonstrated that the fluctuating pair field  results in a
prominent peak of the spectral function, which survives in the
temperature interval $|T-T_c|\lsim (0.1\div 0.2)T_c$. 
Besides, it is
interesting to investigate the role of the order parameter
fluctuations for $T<T_c$, if $T_c$ is $\sim 
50$~MeV, that still may affect the neutrino radiation of the most
massive hot neutron stars (if they indeed 
have quark cores) and the heat transport
at an initial stage of their evolution.

This paper is an extended version of the work \cite{V03}.
We will study precursor phenomena of the CSC
in the framework of the phenomenological Ginzburg - Landau
approach. To be specific, we will consider condensates with 
total angular momentum $J=0$ that are antisymmetric in color and
flavor. Such pairing states can occur in the weak coupling limit
because one-gluon exchange is attractive in the color anti-triplet
channel. Although
for reasonable values of densities and temperatures 
conditions of the applicability
of the weak coupling limit
are hardly fulfilled, in order to get a feeling on possible relevance 
of fluctuation effects
we will still use this limit. At the same time we will by hand vary
the relation $\Delta (\mu_q)$, since the corresponding exponential dependence
is most sensitive to corrections of the running QCD coupling constant
compared to its perturbative value.
We will restrict our discussion by
the consideration of phases, which feasibly 
have large gaps, like 2SC and CFL, as
the most interesting case for applications.

\section*{Physics of  pairing fluctuations}

Like there always exists a vapor under the water, there are excitations on the
ground of any condensate. They appear due to quantum and thermal fluctuations.
In classical systems and also at not too small temperatures 
in quantum systems, quantum fluctuations are suppressed compared
to thermal fluctuations.
Excitations are produced and dissolved with the time passage, although
the mean number of them is fixed
at given temperature.  
Pairing fluctuations are associated with formation and breaking of
excitations of a particular type, Cooper pairs out of the condensate. 
Fluctuation theory of phase transitions is a well developed field.
In particular, 
ten thousands of papers in condensed matter physics
are devoted to the study of pairing fluctuations.
At this instant 
we refer to an excellent review of Larkin and Varlamov \cite{LV}.

In some phenomena pairing fluctuations
behave similarly to quasiparticles.
However there are also differences: 
\begin{itemize}
\item
Typical ``binding energy'' $E_{\rm bind}$ of pairing fluctuation
is of the same order 
as inverse life-time $\tau_{l.t.}^{-1}\propto 1/|T-T_c|$, whereas  
for quasiparticles $E \gg 
\tau_{l.t.}^{-1}$.

\item 
Typical size $l\propto 1/\sqrt{|T-T_c|}$ is large for $T$ near $T_c$.
Quasiparticles have a small typical size.

\item 
Fluctuations near  $T_c$ behave as 
classical fields in sense of Rayleigh-Jeans:
3-momentum distribution is  
 $n(p) \sim T/E(p)$ in the vicinity of   $T_c$.
\end{itemize}

We will demonstrate below that the Ginzburg number ($Gi =\Delta T/T_c$),
which determines the broadness of the energy region  
near the critical temperature, where fluctuations essentially
contribute, is $Gi \sim A (T_c /\mu_q)^4$
with $A\sim 500$ in our case.
To compare, 
for clean metals $A\sim 100$, $\mu_q \rightarrow
\mu_e$, the latter is the electron chemical potential.
Thus $Gi \sim 1$, if $T_c$ is rather high, $T_c \sim
(\frac{1}{3}\div \frac{1}{5}) \mu_q$, 
and we expect  a broad region of temperatures, where
fluctuation effects might be important. 

\section*{Thermodynamical potential
and its mean field solution}

The  Fourier component of  the density of the thermodynamic
potential (thermodynamic potential per unit volume $V$) in the
superconducting quark matter with the di-quark pairing  can be
written in the following form \cite{IB,GR}, cf. also \cite{bl,bss},
\begin{eqnarray}  \label{fe}
\widetilde{\Omega}&=&\widetilde{\Omega}_n + \sum_{\alpha
,i}(-c_0|\partial_{\tau} d_{\alpha}^{i}|^2 +c |\nabla
d_{\alpha}^{i}|^2) +a D + \frac{b}{2} D^2 ~,\,\,\,\,
\\
D &=&\sum_{\alpha ,i}|d_{\alpha}^{i} |^2 ,\,\,\, \gamma =
\frac{1}{D^2}\sum_{\alpha , \beta ,i} |d_{\alpha}^{i, *}\cdot
d_{\beta}^i |^2 ,\,\,\, b=b_1 +\gamma b_2\nonumber \,.
\end{eqnarray}
The Greek indices $\alpha ,\beta =\{R,B,G\}$ count colors, the
Latin indices $i =\{u,d,s\}$ count flavors. The expansion is presented
up to the fourth order in the di-quark field operators (related to the gap)  
assuming the
second order phase transition, although at zero temperature 
the transition might be of the first order, cf. \cite{PR}.
$\widetilde{\Omega}_n$ is the
density of the thermodynamic potential of the normal state. 
The
order parameter squared is $D=|\vec{d}_{\rm IS}|^2 =|\vec{d}_R|^2
+|\vec{d}_G|^2 +|\vec{d}_B|^2$, $\vec{d}_R \| \vec{d}_G \|
\vec{d}_B$ for the isoscalar phase (IS), and $D=3|\vec{d}_{\rm
CFL}|^2$, $|\vec{d}_R|^2 =|\vec{d}_G|^2 =|\vec{d}_B|^2
=|\vec{d}_{\rm CFL}|^2$, \,\,$\vec{d}_R^* \cdot \vec{d}_G
=\vec{d}_G^* \cdot \vec{d}_B =\vec{d}_B^* \cdot \vec{d}_R =0$  for
the CFL phase, $\vec{d}_{\alpha} =\{d_{\alpha}^u , d_{\alpha}^d ,
d_{\alpha}^s\}$. The so called 2SC phase is the IS phase with
unpaired quarks of one color (to say $R$).  
The interaction of the di-quark field with fluctuation gluon fields
is usually introduced in the standard way
through the corresponding gauge-shifted full co-variant
derivatives. Recent paper
\cite{GR03} demonstrated that the gluon fluctuations contribute essentially to
thermodynamic quantities only at
temperatures in a 
narrow vicinity of the critical temperature, $\Delta T/T_c  \lsim 0.1$
for relevant values of parameters. In this rather small temperature interval, 
$\Delta T $, gluon field fluctuations change the nature of the phase
transition (from the second to the first order).
As we will show below, the value $\Delta T$ is much less than the
temperature region, where fluctuations of the di-quark order parameter
might be  important. 
Thereby in the
latter discussion
we suppress the gluon fields as well as the discussion of any
peculiarities of this narrow temperature region near $T_c$.

Near the critical
point coefficients of (\ref{fe}) can be expanded in
$t=(T-T_c)/T_c$. In the weak coupling limit they render:
\begin{eqnarray}  \label{fe-coef}
&&a=a_0 \mbox{ln}(T/T_c )\simeq a_0 t,\,\,\, a_0 =
\frac{2\mu_q^2}{\pi^2},\,\,\,\,\\ &&b_1 =b_2 =b/(1+\gamma )
=\frac{7\zeta(3)\mu_q^2}{8\pi^4
T_c^2},\,\,\,\,
 c=\frac{b}{3(1+\gamma)},\,\,\, c_0\nonumber
=3c.
\end{eqnarray}
For the classical (mean) fields $\gamma =1$ in the IS phase and
$\gamma =1/3$ in the CFL phase, cf. \cite{IB}, $\mu_q =\mu_B /3$,
$\mu_B$ is the baryon chemical potential (the contribution of the
strange quark mass is neglected), $\zeta (3)=1.202...$ The
critical temperature $T_c$ is the same for the IS and  the CFL
phases for $m_s \rightarrow 0$. For $m_s \neq 0$, $T_c$ would depend on
$m_s$ that would result in a smaller  $T_c$
for the CFL phase than for the IS phase.

The  value of the order parameter follows by solving
the  equation of motion for the field operators, $\delta \widetilde{\Omega}
/\delta d_{\alpha}^i =0:$
\begin{eqnarray}\label{meq}
-c_0 \partial^2_{\tau} d^i_{\alpha} +  c\Delta d^i_{\alpha} -a
d_{\alpha}^i-b D d_{\alpha}^i =0.
\end{eqnarray}

The  stationary, spatially-homogeneous mean field solution of (\ref{meq})
(without taking
into  account of fluctuations)   is
\begin{eqnarray}  \label{mf}
D_{\rm MF} =-a\Theta (-t)/b,\,\,\, \delta \widetilde{\Omega}_{\rm
MF }=- \frac{a^2}{2b}\Theta (-t),
\end{eqnarray}
where the step function $\Theta (-t)=1$ for $t<0$ and $\Theta
(-t)=0$ for $t>0$.

To be specific, discussing $T<T_c$ we further  consider the CFL
case. For the finite system of a large spherical size $R\gg \xi$,
with $\xi$ having the meaning of the coherence length, we obtain
\begin{eqnarray}\label{mf-sol}
d_{\alpha ,\rm MF}^i = \pm \frac{\delta_{\alpha}^i}{(N_c N_f)^{1/4}}
\sqrt{D_{\rm MF}}\,\Theta (-t)\, \mbox{th}
\left[\frac{(R-r)}{\sqrt{2}\,\xi}\right],\,\,\, \xi
=\sqrt{\frac{c}{|a|}},
\end{eqnarray}
where $N_c =3$ is the number of colors, $N_f =3$ is the number of
flavors and to be specific we assumed the simplest  structure 
$d_{\alpha ,\rm MF}^i \propto \delta_{\alpha}^i$.

\section*{Fluctuations of gap in self-consistent Hartree approximation}

Now we will consider  fluctuations. Below $T_c$ we present
$d^i_\alpha$, as $d^i_\alpha =d_{\alpha ,c}^i
+d^{i,\,\prime}_{\alpha}$, and above $T_c$, as
$d_{\alpha}^i=d^{i,\,\prime}_{\alpha}$, where index "c" labels the
classical solution.

Since $\delta \widetilde{F} (V,T)=\delta
\widetilde{\Omega} (\mu , T),$ cf. \cite{LL}, the density of the
free energy  of the CFL phase expanded in
$(d^{i,\,\prime}_{\alpha })^2$ terms 
renders
\begin{eqnarray}\label{fr-F}
\delta \widetilde{F}&=& \delta \widetilde{F}_c +\delta
\widetilde{F}^{\,\prime}= \sum_{\alpha ,i}(-c_0|\partial_{\tau}
d_{\alpha ,\,c }^{i}|^2 +c |\nabla d_{\alpha ,\,c}^{i}|^2) +a D_c\nonumber
\\
&+&\frac{b_1 +b_2 /3}{2} D^2_c
+\frac{5}{3}(b_1 +b_2 /3)D_c N_c N_f \sum_{k}
|\phi_k^{\,\prime}|^2 \nonumber \\
&+&N_c N_f \sum_{k} \left( -c_0 \omega^2
+c\vec{k}^{\,2} +a \right)|\phi_k^{\,\prime}|^2 \nonumber \\
&+&\nu_H \frac{b_1 +b_2 }{2}[N_c N_f \sum_{k}
|\phi_k^{\,\prime}|^2 ]^2 
.
\end{eqnarray}
The last term is introduced within the 
self-consistent Hartree approximation 
(within the $\Phi$ functional up to one vertex), $\nu_H =10/9$ accounts
different coefficients in $\Phi$ functional for the self-interaction
terms ($d^4$ - for the given field $d$ 
and $(d^i_{\alpha})^2 (d^j_{\beta})^2$ terms), cf
\cite{IKV}. 
We presented 
$d^{i,\prime}_{\alpha}=\sum_{k}d^{i,\,\prime}_{\alpha
,k}e^{-ik^\mu x_\mu}$, $k^\mu =(\omega , \vec{k})$, $x_\mu =({\tau},
-\vec{r})$, and introduced the notation
\begin{eqnarray}\label{phi-fl}
\sum_{i,\alpha ,k}|d_{\alpha ,k}^{i,\prime}|^2 =\sum_{k}
|\phi_k^{\,\prime}|^2 
=i\int \frac{d^4 k}{(2\pi)^4}\frac{1}{c_0
\omega^2 -c\vec{k}^{\,2} -m^2}.
\end{eqnarray}
In Matsubara technique the temperature dependence is introduced
by the replacements: $\omega \rightarrow \omega_n =2\pi i nT$,
$-i\int \frac{d^4
k}{(2\pi)^4}\rightarrow T\sum_{n=-\infty}^{n=\infty}\int \frac{d^3
k}{(2\pi)^3}$.
We restricted ourselves by the self-consistent Hartree approximation.
Effects of the
damping of fluctuations, being produced by the
two- and more vertex diagrams of $\Phi$, are beyond the scope
of this simple approximation. As we argue below, the main 
effects we discuss in
this paper are not essentially affected by the width terms.

Variation of eq. (\ref{fr-F}) over $\phi_k^{\,\prime}$
yields 
the spectrum of fluctuations
\begin{eqnarray}\label{eq-mfl}
c_0 \omega^2 -c\vec{k}^{\,2} -m^2 =0,
\end{eqnarray}
with the squared mass parameter
\begin{eqnarray}\label{mas} m^2 =\eta |a|=\frac{5}{3}(b_1 +b_2 /3
)D_c  +a+ \nu_H (b_1 +b_2 ) N_c N_f \sum_{k}
|\phi_k^{\,\prime}|^2 \,.
\end{eqnarray}
Notice that the real physical meaning of the
effective mass has the quantity
$m /\sqrt{c_0}$ rather than $m$, as follows from (\ref{eq-mfl}).

Solution of the equation for the fluctuating field presented in the
coordinate space is
characterized by the length scale $l=\xi/\sqrt{\eta }$ and by the
time scale $\widetilde{\tau} =\xi c_0^{1/2} (c\eta )^{-1/2}$.  Above the
critical point $D_c =0$, $ m^2 = a+O(\sum_{k}
|\phi_k^{\,\prime}|^2) >0$, and neglecting $|\phi_k^{\,\prime}|^2$ terms
one gets the parameter $\eta \simeq 1$.

Variation of (\ref{fr-F}) over $d^{i,*}_{\alpha ,c}$
yields the equation of motion for the classical field
\begin{eqnarray}\label{eq-mfcl}
&&-c_0
\partial^2_{\tau} d^{i}_{\alpha ,c} + c\Delta d^{i}_{\alpha ,c}
-\left(a+\frac{5}{3}(b_1 +b_2 /3)N_c N_f \sum_{k}
|\phi_k^{\,\prime}|^2\right)d^{i}_{\alpha ,c}\nonumber \\ 
&&-(b_1 +b_2 /3 )D_c
d^{i}_{\alpha ,c} =0.
\end{eqnarray}
Only dropping the term responsible for fluctuations we find $D_c
=D_{\rm MF}$, $d_{\alpha ,\,c}^i =d^i_{\alpha , \,\rm MF}$, $m^2
=m_{\rm MF}^2 =\eta_{\rm MF}|a|$ for $T<T_c$, cf. (\ref{mf}),
(\ref{mas}), and we obtain $\eta_{\rm MF} =2/3$.
In reality fluctuations affect the classical solution and
renormalize the critical temperature of the phase transition $T_c$
yielding $T_c^{\rm ren} < T_c$.
 More generally
\begin{eqnarray}\label{mren}
m^2 =m^2_{\rm MF}+\delta m^2,\,\,\,\, \delta m^2 =
-\frac{5}{3}(b_1 -b_2 /9)N_c N_f \sum_{k}
|\phi_k^{\,\prime}|^2 ,
\end{eqnarray}
for $T<T_c^{\rm ren}$. From (\ref{eq-mfcl}), (\ref{mren}) one finds
$D_c (m^2 =0)=0$, as the consequence of the self-consistency of our
approximation scheme.
As we have mentioned, we neglected fluctuations of the gluon fields, 
which change the nature of the phase transition but yield  only
a small jump of the order
parameter.

\section*{Contribution of fluctuations of gap to specific heat below $T_c$}

In order to demonstrate how fluctuations of the order parameter may
affect thermodynamic quantities let us calculate the contribution
of fluctuations to the specific heat density $\widetilde{C}_V
=-T\left(\frac{\partial^2 \delta \widetilde{F}} {(\partial
T)^2}\right)_V$, cf. \cite{LL}. The mean field contribution is
\begin{eqnarray}\label{mf-c}
\widetilde{C}_{V}^{\rm MF} =T\frac{a_0^2}{bT_c^{2}} \Theta (-t),
\end{eqnarray}
as it follows from (\ref{fe}), (\ref{fe-coef}) and (\ref{mf}).
Here and below we use an approximate expression $a\simeq a_0 t$
valid at $T$ near $T_c$, see  (\ref{fe-coef}). 
The 
contribution to the specific heat from the normal quark
excitations is suppressed as $\propto T^{-3/2}e^{-\Delta (0)/T}$ 
for $T\ll T_c$, cf. \cite{LP80}. We may reproduce correct low temperature
behavior multiplying
(\ref{mf-c})
by a form-factor 
$f\simeq 2e^{-\Delta (0)/T}(T_c/T)^{5/2}(1-\frac{2T}{3T_c})^{-1}$.

The fluctuation contribution $\widetilde{F}^{\,\prime}$ can be
determined with the help of the functional integration
\begin{eqnarray}\label{efV}
\mbox{exp}(-\delta W) =\int D \phi^{\,\prime} \,\mbox{exp}(-\delta
W[\phi^{\,\prime}]),
\end{eqnarray}
where $\delta W$ is an effective potential. Using (\ref{efV}),
(\ref{fr-F}) we obtain
\begin{eqnarray}\label{fl-f}
\delta \widetilde{F}^{\,\prime} =-iN_c N_f \int \frac{d^4
k}{(2\pi)^4}\mbox{ln}[c_0 \omega^2 -c\vec{k}^{\,2} -m^2].
\end{eqnarray}
Again within the Matsubara technique 
one still should do the replacement $\omega \rightarrow \omega_n =2\pi i nT$,
$-i\int \frac{d^4
k}{(2\pi)^4}\rightarrow T\sum_{n=-\infty}^{n=\infty}\int \frac{d^3
k}{(2\pi)^3}$.
We dropped an infinite constant term in (\ref{fl-f}). However expression
(\ref{fl-f}) still contains a divergent contribution. To remove
the regular term that does not depend on the closeness to
the critical point we  find the temperature derivative of
(\ref{fl-f}) (entropy per unit volume):
\begin{eqnarray}\label{S}
\delta \widetilde{S}^{\,\prime}=\frac{\partial \delta
\widetilde{F}^{\,\prime}}{\partial T} = \frac{\partial
m^2}{\partial T} N_c N_f \sum_{k} |\phi_k^{\,\prime}|^2 .
\end{eqnarray}
The contribution of fluctuations to the specific heat density  is
then given by
\begin{eqnarray}\label{spec-gen}
 \widetilde{C}_V^{\,\prime}&=&
 -T\frac{\partial m^2}{\partial T} N_c N_f \frac{\partial \sum_{k}
|\phi_k^{\,\prime}|^2 }{\partial T}-T \frac{\partial^2
m^2}{\partial
T^2} N_c N_f \sum_{k} |\phi_k^{\,\prime}|^2 \nonumber \\
&\simeq& -T\frac{\partial m^2_{\rm MF}}{\partial T} N_c N_f
\frac{\partial \sum_{k} |\phi_k^{\,\prime}|^2 }{\partial T},\,\,\,
\frac{\partial m^2_{\rm MF}}{\partial T}\simeq -\eta_{\rm MF} a_0
/T_c .
\end{eqnarray}
In the second line (\ref{spec-gen}) we remained only the terms
quadratic in fluctuation fields, i.e. we assumed $|\frac{\partial
m^2_{\rm MF}}{\partial T}|\gg |\frac{\partial \delta m^2}{\partial
T}|$.

Knowing the contribution of
fluctuations to the entropy and specific heat 
we may recover their 
contribution to the free energy and energy densities
\begin{eqnarray}\label{E}
 \delta
\widetilde{F}^{\,\prime}=\int_{0}^{T}\widetilde{S}^{\,\prime}|_{\rho_q} dT,
\quad  \delta
\widetilde{E}^{\,\prime}=\int_{0}^{T}\widetilde{C}_{V}^{\,\prime}|_{\rho_q} dT.
\end{eqnarray}
With (\ref{S}) -- (\ref{E}) one may recover fluctuation contribution to 
all thermodynamic quantities.

Now we are able to calculate the quantity $iG^{- -}(X=0)=\sum_{k}
|\phi_k^{\prime}|^2 $ and its temperature derivative, where $iG^{-
-}(X)$ is the time-ordered Green function. Using the
Matsubara replacement $\omega\rightarrow \omega_n =2\pi i nT$ and
the relation $\sum_{n}(y^2 +n^2)^{-1}=\frac{\pi}{y}\mbox{cth} (\pi
y)$, or the corresponding relation between the non-equilibrium
Green functions $iG^{- -}(X)$ and $\mbox{Im}G^{ret}$ at  finite
temperature, we arrive at the expression
\begin{eqnarray}\label{phiMat}
\sum_{k} |\phi_k^{\,\prime}|^2 =\frac{1}{2\sqrt{c_0}}\int
\frac{d^3 k}{(2\pi)^3}\frac{1}{\sqrt{c\vec{k}^{\,2}
+m^2}}\mbox{cth}\left( \frac{\sqrt{c\vec{k}^{\,2}
+m^2}}{2T\sqrt{c_0}}\right).
\end{eqnarray}
Using that $\mbox{cth}(y/2)=2n_B (y)+1$, where $n_B (y)=(e^y
-1)^{-1}$ are Bose occupations, and dropping the regular
contribution of quantum fluctuations we obtain
\begin{eqnarray}\label{phiMat-1}
\sum_{k} |\phi_k^{\,\prime}|^2_T =\frac{1}{\sqrt{c_0}}\int
\frac{d^3 k}{(2\pi)^3}\frac{1}{\sqrt{c\vec{k}^{\,2}
+m^2}}n_B\left( \frac{\sqrt{c\vec{k}^{\,2}
+m^2}}{T\sqrt{c_0}}\right).
\end{eqnarray}
By index $"T"$ we indicate the thermal contribution.

The integration is performed analytically in the limiting cases. Let
$m\gg T\sqrt{c_0}$. Then $n_B (y) \simeq e^{-y}$. In the very same
approximation one has $c\vec{k}^2 \ll m^2$ for typical momenta.
Then
\begin{eqnarray}\label{phi-it}
\sum_{k} |\phi_k^{\,\prime}|^2_T
=\frac{c_0^{1/4}}{8m}\left(\frac{2mT}{\pi
c}\right)^{3/2}\mbox{exp}\left(-
\frac{m}{T\sqrt{c_0}}\right),\,\,\,  m\gg 2T\sqrt{c_0}\,.
\end{eqnarray}
In the opposite limiting case, $m\ll 2T\sqrt{c_0}$, there are two
contributions to the integral (\ref{phiMat-1}), from the region of
typical momenta $c\vec{k}^{\,2} \sim m^2$ and from the region
$c\vec{k}^{\,2} \gg m^2$. They  can be easily separated, if one
calculates the auxiliary quantity $\frac{\partial \sum_{k}
|\phi_k^{\,\prime}|^2 }{\partial T}$. In the region
$c\vec{k}^{\,2} \sim m^2$ one may use an approximation  $n_B
(y)\simeq 1/y$ and
\begin{eqnarray}\label{phi-itht}
\frac{\partial \sum_{k} |\phi_k^{\,\prime}|^2 }{\partial
T}[c\vec{k}^{\,2} \sim m^2] \simeq -(4\pi c^{3/2})^{-1}
T\frac{\partial m}{\partial T}.
\end{eqnarray}
The region $c\vec{k}^{\,2} \gg m^2$ yields a regular term
\begin{eqnarray}\label{phi-itrel}
\sum_{k} |\phi_k^{\,\prime}|^2  [c\vec{k}^{\,2} \gg m^2]\simeq
\frac{T^2}{12}\sqrt{\frac{c_0}{c^3}},\,\,\,\,\,\, \frac{\partial
\sum_{k} |\phi_k^{\,\prime}|^2 }{\partial T} [c\vec{k}^{\,2} \gg
m^2]\simeq \frac{T}{6}\sqrt{\frac{c_0}{c^3}}\,.
\end{eqnarray}
General expression for the fluctuation contribution to the specific heat
is given by the  first line (\ref{spec-gen}) and can be resolved
with the help of the self-consistent solution of (\ref{mren}), 
(\ref{phiMat-1}) (or (\ref{phi-it}), (\ref{phi-itht}), (\ref{phi-itrel})
in the limiting cases). 
Assuming for rough estimates
that fluctuations can be described perturbatively and
putting $m^2 \simeq m^2_{\rm MF}=\eta_{\rm MF} |a|$, 
from the second line of (\ref{spec-gen})
we find for $T\ll \frac{m_{\rm
MF}}{\sqrt{c_0}}$:
\begin{eqnarray}\label{c-fllt}
\widetilde{C}^{\,\prime}_V &\simeq& \frac{N_c N_f \eta^{7/4}_{\rm
MF}\,\, c^{1/4}\,T^{3/2}}{4\,\pi^{3/2}\sqrt{2} \,c_0^{1/4}\,
|t|^{1/4}T_c^2}\nonumber\\ 
&\times& \left(\frac{a_0}{c}\right)^{7/4}\left(
1+\frac{2|t|T_c}{T}\right) \mbox{exp}\left(-\frac{\sqrt{\eta_{\rm
MF} a_0 |t|}}{T\sqrt{c_0}}\right)
\end{eqnarray}
and for $T\gg \frac{m_{\rm MF}}{\sqrt{c_0}}$:
\begin{eqnarray}\label{c-flht}
\widetilde{C}^{\,\prime}_V \simeq \frac{N_c N_f \eta^{3/2}_{\rm
MF}\,\, T^2}{8\pi T_c^2 |t|^{1/2}}
\left(\frac{a_0}{c}\right)^{3/2}+ \frac{N_c N_f  \eta_{\rm MF}
T^2}{6 T_c } \frac{a_0}{c}\left(\frac{c_0}{c}\right)^{1/2}\,.
\end{eqnarray}
In the weak coupling limit for the CFL phase
$a_0 /c =6\pi^2 T_c^2 /(7\zeta (3)/8)$, $c_0
=3c$, $\eta_{\rm MF} =2/3$. Eq.
(\ref{c-fllt}) holds in the low temperature limit $T\ll \pi
(|t|/3)^{1/2}\,T_c$, whereas eq. (\ref{c-flht}) is valid in the
opposite limit. The first term in (\ref{c-flht})
dominates over the second one 
for $|t| <0.7$, i.e., in the whole region of validity
of (\ref{c-flht}) the main term is the first one.
Its singular
behavior, $\sim 1/\sqrt{|t|}$, is typical for
the specific heat  in the vicinity of the critical point of the
second order phase transition, as in metallic superconductors. 

Also in the CFL phase ($T<T_c$) there is a contribution to the specific heat 
of Goldstone-like
excitations, cf. phonons in the ordinary condensed matter. 
For $T\gg m_{p.G}$, where
$m_{p.G}$ is the mass of the pseudo-Goldstone excitation,
we get
\begin{eqnarray}\label{c-G}
\widetilde{C}^{ p.G, \,\prime}_V =\frac{N_G 3^{3/2} \pi^2 T^3}{30},
\end{eqnarray} 
where $N_G$ is the number of pseudo-Goldstone modes.
One can see that the contribution (\ref{c-G}) is numerically small 
compared to those terms (cf. (\ref{c-fllt}), (\ref{c-flht})) we have
evaluated for temperatures within the fluctuation
region. 

Now we are able to discuss the validity of the self-consistent Hartree
approximation in our problem. Adding the contribution to $\Phi$ functional with
two vertices produces the sun-set diagram in the di-quark self-energy.
In such a way beyond the Hartree approximation
there appears the width term in the self-energy (behaving as 
$-i\gamma \omega$ for small $\omega$). Such a 
term governs the slow relaxation of the 
order parameter in the  phase transition phenomena outside the equilibrium.
In our case (thermal equilibrium)
this term can be dropped compared to the $c_0\omega^2$ term,
which we have in the thermodynamic potential (\ref{fe}) from the very
beginning,
at least in both the low and high temperature limits. 
Indeed, in the low temperature limit three Green functions entering the
sun-set diagram produce an extra exponentially small particle occupation
factor compared to that governs the Hartree term (\ref{phi-it}).
Near $T_c$, i.e. in the high temperature limit, 
we may remain only $n=0$ term in
the Matsubara sum over frequencies, as we have argued,  
thus suppressing both the
linear and the quadratic  terms in $\omega$.

\section*{Ginzburg -- Levanyuk criterion and Ginzburg number}

Comparing the mean field (\ref{mf-c}) and the fluctuation
(\ref{spec-gen}) contributions to the specific heat (in the low
and high temperature  limiting cases  one may use eqs.
(\ref{c-fllt}), (\ref{c-flht})) we may estimate the 
fluctuation temperature $T_{\rm fl,<}^{C}<T_c$, at which the
contribution of fluctuations of the order parameter becomes to be
as important as the mean field one (so called 
Ginzburg - Levanyuk criterion),
\begin{eqnarray}\label{fl-mean}
\widetilde{C}^{\,\prime}_V \simeq \widetilde{C}_V^{\rm MF},\,\,\,
{\rm for}\,\,\, T<T_c .
\end{eqnarray}
Fluctuations dominate for $T>T_{\rm fl,<}^{C}$. For typical values
$\mu_q \sim (350\div 500)$~MeV and for $T_c \gsim (50\div 70)$~MeV
in the weak coupling limit from
(\ref{fl-mean}), (\ref{c-fllt}) we estimate $T_{\rm fl,<}^{C}\simeq (0.6\div
0.8)T_c$.  If we took into account the suppression factor $f$
of the mean field term $\propto e^{-\Delta (0)/T}$, a decrease
of the mass $m$ due to the fluctuation contribution  (cf. (\ref{mren})),
and the pseudo-Goldstone contribution (\ref{c-G}),
we would get
still smaller value of $T_{\rm fl,<}^{C}$ ($\lsim 0.5T_c$).
We see that fluctuations start to contribute at
temperatures when one can still use approximate expressions
(\ref{c-fllt}), (\ref{phi-it}) 
valid in the low temperature limit. Thus the time
(frequency) dependence of the fluctuating fields is important in
case of CSC.

In the condensed matter physics  one usually performs calculations
in the high temperature limit. In this limit one neglects the time
(frequency) dependent terms considering quasi-static 
thermal fluctuations of the order
parameter. Then the fluctuation
contribution is determined with the help of the functional
integration
\begin{eqnarray}\label{fl-o}
\mbox{exp}(-\delta F^{\,\prime}/T) &=&\int D d^{\,\prime}
\,\mbox{exp}(-\delta F^{\,\prime}[d^{\,\prime}]/T),\,\,\,\, \\
\delta
F^{\,\prime}[d^{\,\prime}]&=& \sum_{i, \alpha ,
\vec{k}}(c\vec{k}^{\,2} +\eta |a|)(d_{\alpha
,\,\vec{k}}^{i,\,\prime} )^2 .\nonumber
\end{eqnarray}
The integration yields
\begin{eqnarray}\label{fl-o1} \delta
\widetilde{F}^{\,\prime}= TN_f N_c\int \frac{d^3 k }{(2\pi
)^3}\mbox{ln}[c\vec{k}^{\,2} +\eta |a|].
\end{eqnarray}
Comparison of eqs. (\ref{fl-o1}) and  (\ref{fl-f}) shows that the
high temperature expression (\ref{fl-o1}) is recovered, if
one drops all the terms except $n=0$ in the corresponding
Matsubara sum over $\omega_n =2\pi i nT$ in eq. (\ref{fl-f}). The
contribution of $n\neq 0$ terms to the $\sum_{k}
|\phi_k^{\,\prime}|^2$ is suppressed in the limit $c_0 4\pi^2 T^2
\gg m^2$. Then one immediately arrives at the specific heat given
by the first term of (\ref{c-flht}). Thus in our case one may
suppress the frequency dependence of fluctuations only for $|t|\ll
1$.

Simplifying, the energy width of the fluctuation region, where the
fluctuation effects prevail over the mean field ones, is usually
estimated following the  Ginzburg criterion. The probability of
the fluctuation in the volume $V_{\rm fl}$ is given by $W\sim
\mbox{exp}(-\delta \Omega (V_{\rm fl}) /T)$. It is $\sim 1$
for $\delta \Omega (V_{\rm fl}) \sim T$, where $\delta \Omega
(V_{\rm fl})$ (the contribution to the thermodynamic potential in
the corresponding variables) is the work necessary to prepare the
fluctuation within the volume $V_{\rm fl}$. The minimal size of
the fluctuation region characterized by an order
parameter $d \sim \sqrt{D_{\rm MF}}$ is $\xi(T)\sqrt{2/\eta}$, cf.
(\ref{mf-sol}), (\ref{eq-mfl}). Thus, taking $\delta \Omega
(V_{\rm fl})=\delta \Omega_{\rm MF} (V_{\rm fl})= T_{\rm fl}^{\rm
G}$ for the typical temperature $T_{\rm fl}^{\rm G}$, when
fluctuations start to dominate, we obtain
\begin{eqnarray}\label{Ginz}
T_{\rm fl}^{\rm G}\simeq \frac{a^2}{2b}\frac{4\pi (\sqrt{2/\eta}\,
\xi (T_{\rm fl}^{\rm G}))^3}{3}\simeq \frac{ 4\pi a_0^{1/2}c^{3/2}
|t(T_{\rm fl}^{\rm G})|^{1/2}}{\eta^{3/2}b} \frac{\sqrt{2}}{3}.
\end{eqnarray}
Although the above estimate is very rough we took care of all
the numerical factors. This allows us to notice that the  value
$T_{{\rm fl},<}^{C}$ estimated from (\ref{fl-mean}), if one uses
eq. (\ref{c-flht}) for $\widetilde{C}^{\,\prime}_V $ remaining
there only the first term, is $T_{{\rm fl},<}^{C}\simeq 0.5
|t(T_{{\rm fl},<}^{C})/t(T_{\rm fl}^{\rm G})|^{1/2} T_{\rm
fl}^{\rm G}$ for $N_c =N_f =3$. All the dependencies on the
parameters in expressions for $T_{{\rm fl},<}^{C}$ and
$T_{\rm fl}^{\rm G}$ were proven to be essentially the same.

Fluctuation region is determined by the Ginzburg number
$Gi =|T_c -T_{\rm fl}^{\rm C}|/T_c$. 
The larger the value $Gi$ is the broader is the
fluctuation region. For clean conventional superconductors
\cite{LV} $Gi \simeq A(T_c /\mu )^4
\sim 10^{-12}\div
10^{-14}$, $A\sim 80$,
whereas for  superfluid He$^4$ and in our case $Gi \sim 1$ (since
$A \sim 500$
and $T/\mu_q$ can be as large as $\frac{1}{3}\div \frac{1}{5}$ 
in favorable cases).
The role of fluctuations increases in cases, 
when the effective dimensionality of the 
droplet decreases or/and when 
the quark mean free path decreases. Both possibilities 
result in a significant
increase of the $Gi$-number.
In the case of
quark droplets of the typical size $L\ll \xi$
one deals with the 
zero-dimensional
system, $Gi \propto \xi^3 /L^3$
and the fluctuation contribution to the specific heat 
behaves as
$\widetilde{C}_V^{\prime} \sim 1/t^2$, for $Gi\ll |t|\ll 1$. 
For dirty superconductors $Gi$ increases $\propto 1/(p_{Fe} l_e)$, where 
$p_{Fe}$ is the electron Fermi momentum and $l_e$
is the electron mean free path.
A significant decrease of the quark mean free path is expected
due to the presence of the hadron impurities inside the quark droplet.
Thus, there is still a variety of possibilities for a further enhancement
of superconducting fluctuations.

\section*{Fluctuations of the gap above $T_c$} 

All above expressions
for fluctuating quantities (except the vanishing of the
pseudo-Goldstone contribution above $T_c$ and the appearance of an extra
di-quark decay contribution to the width) are also
valid for $T>T_c$, if one puts $D_c =0$ in general expressions
or $m^2_{\rm MF} =a>0$, $D_c =0$,  $\eta
=\eta_{\rm MF}=1$ in the corresponding approximate 
expressions.
Above $T_c$ we
should compare the fluctuation contribution to a thermodynamic
quantity with the quark and gluon contributions of the normal
state of the quark-gluon plasma. In the weak coupling limit for
the specific heat density of the quark-gluon plasma one has, cf.
\cite{Mul},
\begin{eqnarray}\label{norm-c}
\widetilde{C}^{qg}_V \simeq 6 \mu_q^2 T  +\frac{42\pi^2
T^3}{15}+\frac{32\pi^2 T^3}{15}.
\end{eqnarray}
The first two terms are quark contributions and the third term is the
gluon contribution. We omitted a contribution of strange quarks
which is rather small for $T<m_s$ and we neglected the temperature
dependent effective gluon mass. We also disregarded the $\alpha_s$
corrections to the quark and gluon terms since such corrections
were not taken into account for condensate quantities.
If we included the quark-gluon masses that appear in the framework of the
approach \cite{P} matching the quasiparticle quark-gluon description 
and the lattice results, we would get even smaller contribution of 
$\widetilde{C}^{qg}_V$, that is in favor of fluctuation effects.

Above $T_c$, eqs. (\ref{c-fllt}) and (\ref{c-flht}) yield rather
smooth functions of $T$ except the region $t\ll 1$. The fluctuation
region is rather wide since $\widetilde{C}^{\,\prime}_V \sim
\widetilde{C}^{qg}_V$ even at $T$ essentially larger then  $T_c$.
The appearance of an extra channel of the di-quark 
decay width, $2\gamma_{\rm dec}\omega$,
beyond the Hartree approximation
does not qualitatively change the situation since $\gamma_{\rm dec}$
is a regular function of $T$,
$\gamma_{\rm dec}\propto T-T_c$, and vanishes in the critical point. Thus
the applicability of the high temperature limit 
(one can put $\omega_n =0$ in the
Matsubara sum) is preserved in a wide temperature region, $T\sim T_c$, 
\cite{LV}. 
Only far above $T_c$ situation might be changed.
The main uncertainty comes from the values of  coefficients (\ref{fe-coef}) 
which were derived for
$T$ rather 
near  $T_c$ and in the weak coupling limit. Bearing all this in mind 
we estimate the value of the fluctuation temperature
$T_{{\rm fl},>}^C \sim 2T_c$ for typical value $T_c /\mu_q \sim
0.1\div 0.3$. The higher this ratio is, the stronger is the
contribution of fluctuations at the given ratio $T/T_c$.
Note that a decrease of $\mu_q$ with increase of the temperature
results in an increase of the ratio $T_c /\mu_q (T)$. On the other
hand $\mu_q$ should be at least larger than zero temperature gap in order one
could speak of any pairing fluctuations at given temperature..

Thus we see that {\em fluctuations of the di-quark gap may essentially
contribute to the thermodynamic quantities even well below $T_c$
and  above  $T_c$.}

\section*{Assumptions which we have done}

Note that several simplifying  assumptions were done. The
coefficients
(\ref{fe-coef}) were derived for $|t|\ll 1$ in the weak coupling
limit  neglecting the strange quark mass, but applied 
in a wider temperature region for may be not sufficiently large $\mu_q$.  
Only Gaussian
fluctuations were taken into account within the self-consistent 
Hartree approximation.  Thus,  di-quark width effects were assumed to be
suppressed. 
We used the expansion of the
thermodynamic potential up to the fourth order terms in the mean
field neglecting a small jump in the order parameter 
due to gluon fluctuations. As was argued in \cite{GR03}
gluon fluctuations
may also contribute but 
in a more narrow vicinity of $T_c$ 
than the order parameter fluctuations (for the values of the
ratio $T_c /\mu_q \sim 0.15\div 0.3$, which we are interested here). 
We incorporated in the thermodynamic quantities
only the terms which have a tendency to 
an irregular
behavior near $T_c$, whereas above $T_c$ the short range 
correlations begin to be more and more important with the increase of the
temperature. 
Therefore one certainly should be cautious applying
above rough estimates outside the region of their quantitative
validity. However, as we know from the experience of the
condensate matter physics, see \cite{LV}, such extrapolation equations work
usually not too bad even for temperatures well below and above
$T_c$.

\section*{Fluctuations of temperature, density, magnetic susceptibility}

So far we have discussed the specific behavior of fluctuations of
the order parameter at fixed temperature and density.
There are also fluctuations of the temperature and the local quark
density.  They are statistically independent quantities \cite{LL},
$<\delta T\delta\rho_q
>=0$, and their mean squares are
\begin{eqnarray}\label{fl-tr}
<(\delta T)^2 > =\frac{T^2}{V_{\rm fl} \widetilde{C}_V },\,\,\,
<(\delta \rho_q)^2 > =\frac{T\rho_q}{V_{\rm
fl}}\left(\frac{\partial \rho_q} {\partial P}\right)_T.
\end{eqnarray}
The averaging is done over the volume, $P$ is the pressure,
$V_{\rm fl}$ is as above the volume related to the fluctuation. In
the limiting case $\widetilde{C}^{\,\prime}_V \ll
\widetilde{C}^{qg}_V$, we obtain $<(\delta T)^2 >/T^2 \simeq
\rho_q /(N^{\rm fl}_q \widetilde{C}_V^{qg})$, where $N^{\rm fl}_q$
is the number of quarks involved in the volume $V_{\rm fl}$. Thus,
far from the critical point the contribution of fluctuations is
suppressed as $\sqrt{<(\delta T)^2
>/T^2}\propto 1/\sqrt{N^{\rm fl}_q}$. This standard fluctuation
behavior is essentially changed for $\widetilde{C}^{\,\prime}_V
\gsim \widetilde{C}^{qg}_V$. Then we get even larger suppression
of the temperature fluctuations. For $|t|\ll 1$ in the high
temperature limit we obtain $\sqrt{<(\delta T)^2
>/T_c^2} \sim |t|\rightarrow 0$ for
$|t|\rightarrow 0$, where we assumed that fluctuations are most
probable within the typical fluctuation volume $V_{\rm
fl}=4\pi \xi^3 /3$ and $\xi \propto 1/\sqrt{|t|}$ for $t
\rightarrow 0$. Thus, if the system is at the temperature $T$ in a
narrow vicinity of $T_c$, fluctuations of the temperature are
significantly suppressed. Being 
formed at $0< -t\ll 1$ the
condensate region evolves  very slowly, since the heat transport is then
delayed (the typical evolution time is roughly $\tau\propto
1/\sqrt{|t|}$). At temperatures $T$ outside a narrow
vicinity of $T_c$, fluctuations of the temperature resulting in
a significant decrease of the temperature in the volume $V_{\rm fl}$  
are rather probable, $\sqrt{<(\delta T)^2
>/T_c^2} \sim 1$. This is the
consequence of a very short coherence length and, thus, not too
large number of quarks contained in the volume $V_{\rm fl}$, $\xi
\simeq 0.13/(T_c \sqrt{|t|})$ being $\simeq 0.5$~fm for $T_c
\simeq 50$~MeV and for $|t|\sim 1$.  
Fluctuations of the quark density are also large
for typical $|t|\sim 1$, due to the smallness of $V_{\rm fl}$.
Thereby, we  argue that the system may produce
the di-quark condensate regions of the typical size $\xi$,  thus
feeling the possibility of the phase transition even if its
temperature and density are in average rather far  from the
critical values.

Other quantities associated with second derivatives of the
thermodynamic potential are also enhanced near the critical point
demonstrating typical $1/\sqrt{|t|}$ behavior, cf. \cite{LP80}. 
However numerical
coefficients depend strongly on what quantity is
studied.
E.g. fluctuation contributions above $T_c$ to the
color diamagnetic susceptibilities
\begin{eqnarray}\label{fl-g}
\chi_{\alpha} =-\left( \partial^2 \delta F /\partial
{\vec{\cal{H}}}_{\alpha}^2 \right)_{{\vec{\cal{H}}}_{\alpha} =0},\,\,\,
{\vec{\cal{ H}}}_{\alpha} =\mbox{curl}{\vec{\cal{A}}}_{\alpha}\,,
\end{eqnarray}
are proven to be  $ \ll 1$ everywhere except very narrow vicinity
of the critical point. In spite of a smallness, as we know, 
in  metals the
fluctuation diamagnetism turns out to be of the order of the Pauli
paramagnetism even far from the transition. Also 
for $T\gg T_c$ contribution of fluctuations to
the 
magnetoconductivity of 2D electron systems is  experimentally 
distinguishable,  \cite{LV}.

\section*{How gap
fluctuations may manifest in heavy ion collisions}

Anomalous behavior of fluctuations might
manifest itself in the event-by-event analysis of the heavy ion
collision data. In small ($L$) size
systems, $L<\xi$, (zero dimension case would be $L\ll \xi$)
the contribution of fluctuations of the order parameter
to the specific heat is still increased, as we have mentioned, see
\cite{LV}.
The anomalous behavior of the specific heat may affect the heat
transport. Also kinetic coefficients are substantially affected by
fluctuations
due to the shortening of the particle mean free paths, as the consequence 
of their
rescatterings on di-quark fluctuations.
If thermalization happened at an initial heavy ion
collision stage, 
and a large size system
expands rather slowly, its evolution is governed by the
approximately constant value of the entropy. Due to a large
contribution to the specific heat (and to the entropy) of 
di-quark fluctuations, an  extra decrease of the temperature may
occur  
resulting in an essential slowing of the fireball  expansion
process (due to smaller pressure). 
Having the di-quark quantum numbers, fluctuations of
the gap  may affect the di-lepton production rate from the
quark-gluon plasma, as it has been noted in \cite{KKKN02}.
Another question is how 
one can distinguish 
di-quark fluctuations related to the CSC 
from the quark fluctuations, which may relate to the deconfinement
transition? Theoretically, since $T_c$ is, in general, different from
the deconfinement transition temperature
$T_{dec}$, in assumption that $T_c >T_{dec}$
there might be two
anomalous fluctuation peaks related to two distinct collision energies. 
E.g., if the observed peak \cite{Af02} in  the 
$K^+/\pi^+$ ratio  at  $\sim (30\div 40)~$GeV/A
is, indeed, associated with the deconfinement transition
\cite{GG},
it would be worth to seach a
possibility of  another peak at a somewhat higher collision
energy than the first one, related to the
CSC phase transition, and vice versa, if the observed peak relates to
the CSC, another peak at a 
somewhat lower energy could relate to the deconfinement.
However it remains
unclear 
is it possible to distinguish these peaks experimentally,  
if $T_c$ and $T_{dec}$ are rather close to
each other. One needs a careful measurement of  $K^+/\pi^+$ ratio in the whole
collision energy interval.

\section*{Pairing fluctuations in hybrid stars}

Besides the crust and the hadron shell,
the hybrid star contains also a quark core.
Both the nucleon shell and the quark core can be in superconducting phases,
in dependence on the value of the temperature.
Fluctuations affect transport coefficients, specific heat,
emissivity, masses of low-lying excitations and respectively  
electromagnetic properties of the star, like electro-conductivity and magnetic
field structure, e.g., renormalizing 
critical values of the magnetic field ($H_{c1}$,
$H_{c}$, $H_{c2}$). Note that
the CSC is the type I superconductor with the Ginzburg-Landau
parameter $\kappa <0.6$, if $T_c <14~$MeV (estimation is done
for $\mu_q \simeq 400$~MeV),
and it is the type II superconductor with $\kappa $ essentially larger 
than one, if $T_c$ can indeed be large,
$T_c \gsim 50~$MeV, that is the case which we are interested in 
here, cf. \cite{GR03}. Especially the type II CSC could have interesting
consequences for observations, cf. \cite{bss}.

The effect of thermal 
pion fluctuations on the specific heat
and the neutrino emissivity of neutron stars 
was discussed in 
\cite{VS96,SV} together with  other in-medium
effects, see also reviews \cite{MSTV,V00}.
Neutron pair  breaking  and formation (PBF) neutrino process
on the neutral current 
was studied in \cite{FRS,VS87} for the hadron matter. Also 
ref. \cite{VS87} added the proton PBF process in the hadron matter
and correlation processes,
and ref. \cite{JP} included 
quark PBF processes in quark matter.
PBF processes were studied by two different methods;
with the help of Bogolubov 
transformation for the fermion wave function \cite{FRS,JP}
and within  Schwinger-Kadanoff-Baym-Keldysh formalism for non-equilibrium
normal and anomalous fermion Green functions \cite{VS87,SV,MSTV}.

As was observed in \cite{VS87,SV},
analogously to the Direct Urca (DU) process $n\rightarrow pe\bar{\nu}$,
these processes
($n\rightarrow n\nu\bar{\nu}$ and $p\rightarrow p\nu\bar{\nu}$)
have a large (one-nucleon) phase space volume, if the pairing gap is
$\Delta \gsim 1~$MeV in some density interval (in addition to that
DU and PBF emissivities have
similar exponential suppression factors $\propto
e^{-\Delta /T}$). Moreover, these processes are affected by 
nucleon-nucleon \cite{VS87} and electron-electron \cite{VKK,L,V00}
correlations in such a way that the emissivity of the process
$n\rightarrow n\nu\bar{\nu}$ is not significantly changed but
the emissivity of the process 
$p\rightarrow p\nu\bar{\nu}$ increases up to ten -- hundred times.
This enhancement is due to the fact that the square of the bare vertex for the
$p\rightarrow p\nu\bar{\nu}$
reaction contains a very small
$\sim c_v^2 \simeq 0.006$ factor compared to the corresponding factor
$\sim 1$ for  the $n\rightarrow n\nu\bar{\nu}$ reaction 
channel. However the proton  
may produce the neutron-neutron
hole by the strong interaction
and the electron-electron hole by the electromagnetic 
interaction, which then may couple to the weak current. 
This circumstance
is still ignored in a number of 
works, which rediscovered this process
with vacuum vertices and used it within the cooling code,  e.g.,
see \cite{Y}. It seems that an artificial a ten -- hundred times suppression 
of the relevant process  may affect their conclusions.
Numerical 
simulation of the neutron star cooling 
that incorporated PBF processes with inclusion of
correlation effects,
as well as other relevant in-medium effects of the nucleon-nucleon
and nucleon-pion
interaction, like softening of pion modes, was performed in 
\cite{SVSWW}. 
The PBF processes in the quark matter are also affected by correlation
effects but, as for the case of the reaction 
$n\rightarrow n\nu\bar{\nu}$,  in the given case the 
correlation effects are expected to be not
so significant. 

Order parameter fluctuations allow for extra neutrino processes
having no exponential suppression in a broad region of temperatures near
$T_c$.

Contribution of pairing fluctuations to the specific heat in
the hadron shell is minor for the case of the
neutron pairing due to a small value of $T_c \lsim ~1$MeV compared to
the value of the neutron chemical potential ($\mu_n \gsim 50~$ MeV).
Therefore in the neutron channel
fluctuations of the gap are relevant only in a very
narrow vicinity of the critical point.
However this effect might be  not so small for protons,
for which the chemical potential is of the order of several MeV,
whereas the gap is of the order of one MeV. Therefore it seems
that fluctuations may smear
the phase transition in a rather broad vicinity of the critical point 
of the proton superconductivity.

Pairing fluctuation effects  in the quark matter are especially
important, if the critical temperature of the CSC phase transition
is rather high ($T_c \gsim 50~$MeV), as is estimated for 2SC and CFL cases.
Temperatures $\sim 10\div 50~$MeV may indeed 
arise at an initial stage of the hybrid star
cooling. 
The fluctuation 
contribution to the total specific heat
for $T\lsim T_c$ is evaluated in a line with that we have done  above.
As noted in ref. \cite{LV}, the Aslomasov - Larkin contribution to the
electro-conductivity proves to be 
the dominant term in the vicinity of $T_c$.
Analogously, one may expect a significant increase of the heat conductivity
and viscosity
that would result in a delay of the heat transport at the initial  stage
of the hybrid star cooling.
Moreover, neutrino may efficiently rescatter on fluctuations
that governs their drift to the hadron shell at  $T\lsim T_c$.
These effects may be especially important for the CFL phase, since
in that case
all other mechanisms of the heat transport  are suppressed at $T<T_c$.
Estimation  of all  these effects needs however a separate study.

\section*{Concluding remarks}

We would like once more
to emphasize that the coefficients of the thermodynamic
potential (\ref{fe}) obtained in the weak
coupling limit might be essentially modified, in case if we applied the
results to the fireball produced
in a heavy ion collision or to hybrid stars, 
somehow changing our conclusions. E.g.,
the quark chemical potential decreases with the temperature. It results in an
additional increase of the contribution of fluctuations, cf. dependences 
of the coefficients (\ref{fe-coef}) on
$\mu_q$. Thus we may need to analyze
the  strong coupling limit instead of the weak coupling limit we discussed. 
There exist arguments that the strange quark mass $m_s$ is very large
and due to that the phase transition in the CFL phase
does not occur up to very high baryon densities \cite{CKB}.
If $\mu_q$ becomes
smaller than the strange quark mass we come from the possible 3SC phases 
to the 2SC phases. 
A more general discussion of modifications
beyond the framework of the weak coupling limit can be
found in \cite{P00}. A discussion how the
coefficients (\ref{fe}) may in general vary within the 
Ginzburg - Landau approach is given in \cite{IB}.      
Our above arguments are admittedly speculative, and mean only to demonstrate a
qualitative
possibility of their application to heavy ion collisions and to an initial
stage evolution of the hybrid star.
A more detailed treatment of the problem definitely needs a further work.

Concluding, within  the Ginzburg - Landau approach we estimated
the fluctuation energy region at the CSC phase transition. The quantitative
estimates are based on the values of
parameters derived in the weak coupling limit for 
fluctuations of the CFL order parameter. Qualitative results survive
also for fluctuations of other possible phases.
We found that 
the frequency
dependence of fluctuations is important for CSC in a wide
temperature region. Fluctuations may contribute essentially to the
specific heat even at $T$ rather far below and above $T_c$. We
estimated $T_{{\rm fl},< } \lsim 0.5 T_c$ and 
$T_{{\rm fl},> }\sim 2 T_c$. 
Our rough estimates show that the high temperature CSC could
manifest itself through fluctuations of the di-quark gap in the course
of the heavy ion collisions at SIS300, if the critical temperature
of the phase transition is indeed rather high, $T_c \gsim (50\div
70)$~MeV. 
The CSC fluctuations are also relevant for an initial stage of the hybrid star
evolution, if $T_c$ is  $\sim 50~$MeV.
However quantitative results depend on the values of
several not sufficiently known parameters and  further studies are still
needed to arrive at  definite conclusions.


\begin{acknowledgments}
The author thanks D. Blaschke, T. Kunihiro and M.F.M. Lutz
for the discussions. He acknowledges
the hospitality and support of GSI Darmstadt. The work has been
supported in part by DFG (project 436 Rus 113/558/0-2), and by
RFBR  grant NNIO-03-02-04008. The author also
acknowledges NATO Scientific Program for the support of
his participation in Advanced Research Workshop in Erevan, Sept. 2003.
                     
\end{acknowledgments}

\end{document}